# Statistical thermodynamics of the binding energy super-landscape in the adaptive immune system

*József Prechl*

R&D Laboratory, Diagnosticum Zrt., Budapest, Hungary

ORCID: 0000-0003-3859-4353

Correspondence: József Prechl

jprechl@diagnosticum.hu

## Abstract

Adaptive humoral immunity, from the physical perspective, can be regarded as the self-organization of the binding energy landscape of antibodies. In biological terms, the humoral immune system evolves and adapts its repertoire of antigen binding molecules so as to maintain its molecular integrity by controlled removal of antibody-antigen complexes. Here we introduce a super-landscape model, created by the fusion of binding energy landscapes of the antibody repertoire, that can be described by the distribution of interaction energies and deformation parameters of chemical thermodynamic potentials in the system. These deformation parameters not only characterize the partition function of the ensemble and the network of interactions in the system but also the asymmetry of generalized logistic distributions obtained in immunoassays when probing the system. Overall, a statistical thermodynamics approach is provided for a deeper theoretical insight into the dynamical self-organization of the adaptive immune system and into the interpretation of experimental results of immunoassays.

## 1. Introduction: self-organization and thermodynamics of the adaptive immune system

Vertebrate animals possess a complex system of cells and molecules that rivals the central nervous system in numerosity and diversity [1]: the adaptive immune system. While the central nervous system adapts the host animal to its macroscopic physical environment, the adaptive immune system controls the molecular environment by maintaining cells and molecules capable of removing their targets. Adjustment of the efficiency of this removal shapes the landscape of targets and maintains molecular integrity of the host [2]. This is what we perceive as protection against infectious agents and tumor cells, as holding the immense microbiota at bay and as the clearance of cellular waste material. Therefore, maintenance of molecular integrity requires the maintenance of constant concentrations of effector molecules, which are called antibodies (Ab). This is achieved by the adjustment of chemical potentials with the help of a sensor-effector feedback mechanism [3], which is the essence of the phenomenon we call immunity. The immune system is dynamic, continuously responding to environmental stimuli, but also shows a tendency to come to "rest", contract and reach a thermodynamically optimized steady state [4], where minimal effort is required for its maintenance. The system is embedded in a thermodynamic reservoir, the host organism, which maintains constant temperature and regulates chemical potentials by removing complexes of Ab and bound antigen (Ag). An immune response is triggered by increased antigen chemical potential, which is detected by sensor B cells [5] (Fig.1). The system then amends Ab chemical potential and drives the flow of AbAg complexes across the system border, thereby readjusting antigen levels.

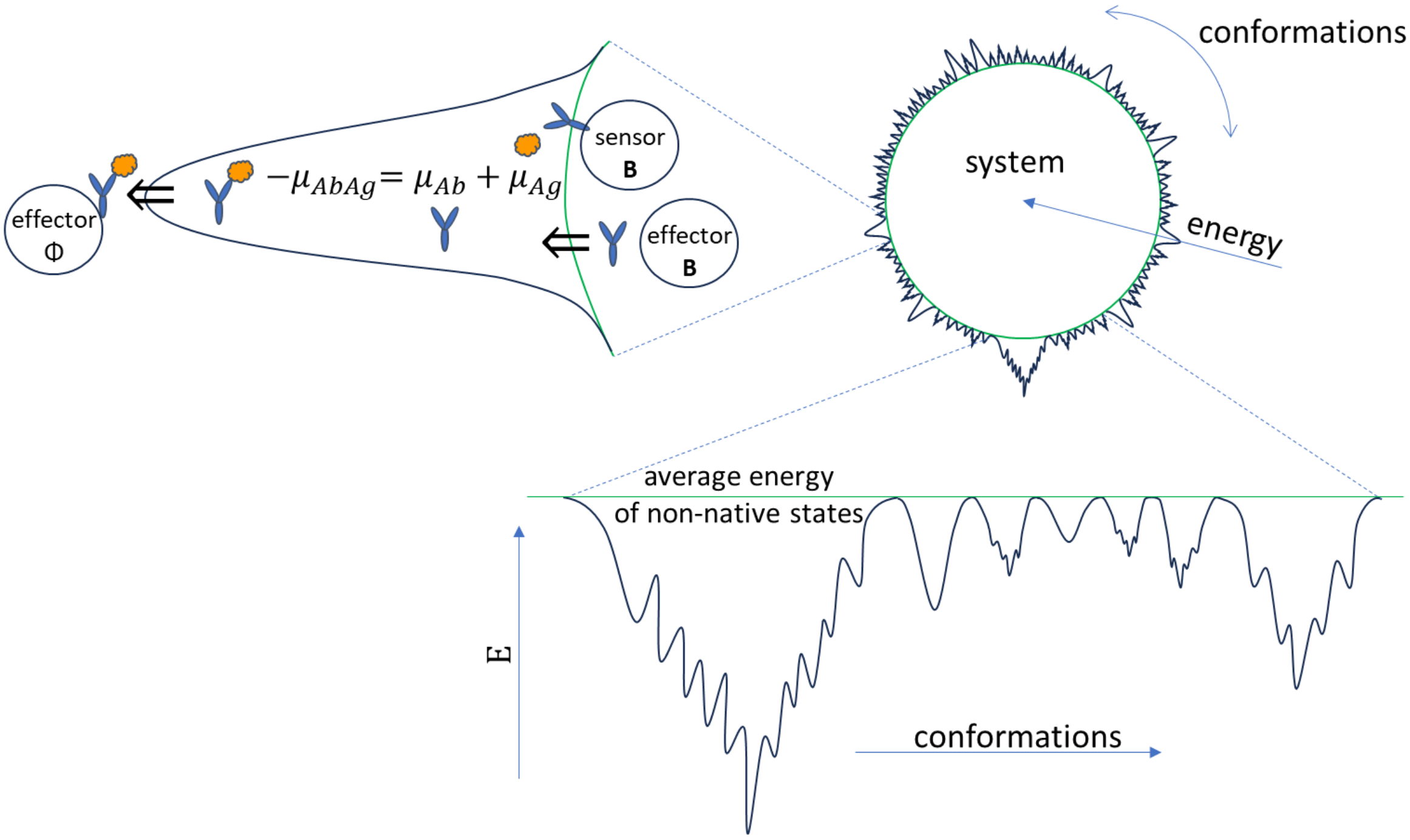

*Figure 1. Energy super-landscape and chemical potentials in the humoral immune system.*
*The binding energy super-landscape forms a thermodynamic system with multiple components. For each binding funnel chemical potentials are adjusted by cells of the immune system (circles), which sense Ag abundance (sensor B lymphocyte), secrete Ab molecules (effector B lymphocyte) and remove Ag-bound Ab (effector phagocytic cell). Cellular differentiation can adjust binding affinity as required. Steady state non-equilibrium is maintained by the flow of antigens through the system. B, B lymphocyte; Φ, phagocyte*

The Ab responses can include both the tuning the strength of non-covalent binding (a process called affinity maturation) of Abs and the change of Ab concentration, corresponding to enthalpic and entropic contributions to Ab chemical potentials, respectively. Chains and networks of interactions shape the landscape of Ab chemical potentials as a result of overlapping conformational landscapes [3,6,7]. These events can be monitored by biological techniques that assess the breadth and depth of the immune repertoire on the level of protein sequences [8–13], and various models are used for the analysis and interpretation of the observations [14–18]. Few models exist however that employ universal, statistical physical approaches to the system [19,20]. In the following sections we examine how statistical distributions that are conventionally used in thermodynamics can be applied to and interpreted in the description of adaptive immunity and in the analysis of experimental measurements.

## 2. Funnel energy landscapes of antibody binding

Statistical mechanics and energy landscapes were originally introduced for the modeling of protein folding [21,22]. A funnel shaped energy landscape that guides molecules from conformational diversity towards thermodynamic stability not only helped visualize entropy-energy compensation in the process of folding but generated answers about the thermodynamics, kinetics and evolution of macromolecules and their interactions [23–29]. It turns out that binding mechanisms, where intermolecular interactions supplement intramolecular interactions, can also be explained by funnel energy landscapes [23] and free energy landscapes in general [30,31]. It is therefore reasonable to apply this model to a biological system, which regulates extracellular molecular interactions: humoral immunity – primarily but not exclusively – adjusts the concentrations of target molecules, Ags, via the directed evolution of a system of Ag binding proteins, the Abs. Here, we assume that immunological self-organization drives the system of antigen and antibody molecules towards a steady state, which encompasses the fusion of binding energy landscapes of individual antigens and antibodies, generating a super-landscape (Fig.1).

We regard the totality of interacting Ag and Ab molecules as an ensemble of conformational isomers, with conformational diversity originating both from protein sequence differences (molecular or clonal diversity) and structural dynamism (conformer diversity). That antibody conformational isomerism can contribute to effective structural diversity [32,33] and is modulated by antibody maturation [34] has long been recognized. While individual Abs have been treated as conformational ensembles of the binding site [35], and of the Ag binding fragment [36,37], the modeling of the complete repertoire as an ensemble of fused binding energy landscape of dynamic conformational ensembles holds the promise of a physical model of the humoral immune system. Indeed, landscapes of Ab-Ag interactions are recently being used to characterize immunity [38–40].

In the energy funnel model of binding the free energy of binding is given by the equation [27,41,42]

$$\Delta G = E_N + \frac{1}{\beta} ln\left[\sum_{E>E_N} g(E) e^{-\beta E}\right] \quad (1)$$

where ΔG is free energy difference, $E_N$ is the ground-state energy of the native structure, β is thermodynamic β=1/kT (k is Boltzmann constant, T is thermodynamic temperature), g(E) is the density of states, E is energy level.

Let us apply this model to the interaction of an Ag molecule with serum Abs. The quality of interacting Abs determines the level of the lowest energy state $E_N$ of the Ag molecule. The quantity of interacting Abs determines the distribution of Ag molecules above this energy level. Thus, equation (1) tells us that the free energy gradient sustained by the immune system is determined by energy of the native state (first part of sum) and the thermodynamic states of molecules in the funnel (second part of sum). Immunological mechanisms adjust both funnel depth and breadth: immunogenic Ag drives antibody maturation leading to increased affinity (decreased $E_N$), while excess antibody secretion modulates the density of states in the funnel (Fig.1). Immunological self-organization can therefore be described as the shaping of the antigen energy landscape: moving antigen molecules deemed dangerous by the immune system to deeper funnels, with increased stability of their bound forms. While the quality of the antibodies determines the depth of the funnels, the quantity and cross-reactivity determines how deep Ag molecules are driven into the funnels – events that can be modeled by physics theory.

## 3. Properties of the super-landscape of serum antibodies

In our model we assume that Ag binding energy of serum Abs is exponentially distributed, based on experimental and theoretical reasons. Models of fluctuating antigenic landscape [43] and experimental determination of clone sizes [44] revealed that lymphocyte clone sizes follow power law. Power law distribution is generated when deterministic exponential growth is stopped at random time, which is exponentially distributed [45]. It follows that if antigen stimulus induces exponential growth of lymphocytes and is stopped at exponentially distributed time intervals, clone size is distributed according to power law. Antigen stimulation for exponentially distributed time intervals can result in an exponential distribution of antigen binding energies in the system via sustained affinity maturation, the selection of B-cell clones with the appropriate affinity and their differentiation into Ab secreting plasma cells.

Let the energy states of Ab-Ag molecule complexes in a funnel energy landscape of binding be distributed according to

$$p(\mathrm{E}) \propto e^{-\beta\mathrm{E}} \qquad (2).$$

We can introduce a factor, $\nu_H$, to account for the changes in binding energy distribution resulting from the fusion of all the individual binding Ab energy landscapes into the funnel. Then the energies are distributed according to

$$p(\mathrm{E}) \propto e^{-\frac{1}{\nu_H}\beta\mathrm{E}} \qquad (3)$$

if the expected value of energies in the unfused landscape is a $\nu_H$-th fraction of the expected value of actual energy states. In other words, $\nu_H$ is a proportionality factor between the enthalpic contribution of Ag and the excess enthalpy contributed by Ab molecules to the binding ensemble (Fig.2).

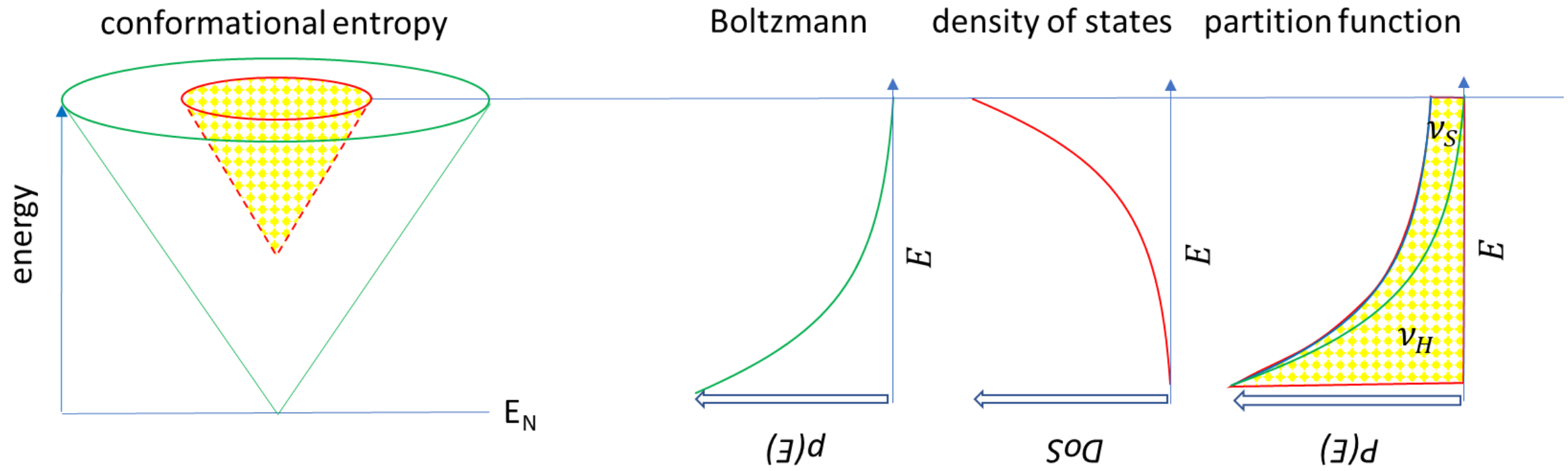

*Figure 2. Distributions of conformational states in the binding energy landscape*
*Statistical distributions of conformational microstates can be modeled by the Boltzmann distribution and the density of states. Because of the gradient maintained by constant Ab generation and Ab-Ag complex removal shown in figure 1, we model the density of states with an exponential curve (see details in text). The area under the curve of the product of these distributions is the partition function of the system. Asymmetry parameters that are rate parameters in the exponential functions represent proportionalities in the partition function.*

In the fused super-landscape, we can express the density of states by introducing another factor, $\nu_S$, that represents the relationship between configurational entropy directly associated with Ag conformation and the conformational entropy added by the fused Ab funnels (Fig.2). We assume an exponential increase of conformational space with energy, as this is a regime of flexible epitope-paratope interactions where ensembles of single clones and clonal diversity contribute to conformational entropy. Although the density of states is an exponential of entropy [46], the actual number of accessible conformations is limited by the concentration of serum antibodies that constantly populate the funnel (Fig.1). So, we model the density of states g(E) with an exponential function that relates the number of accessible conformational states to the number of states dictated by a strict enthalpy-entropy compensation in the funnel, using the deformation factor $\nu_S$ as exponential rate: $e^{\nu_S \beta \mathrm{E}}$. Now we obtain the partition function of the ensemble by integration

$$\mathrm{Z} = \int e^{\nu_S \beta \mathrm{E}} e^{-\frac{1}{\nu_H}\beta \mathrm{E}} d\mathrm{E} = \int e^{-(\frac{1}{\nu_H}-\nu_S)\beta \mathrm{E}} d\mathrm{E} \quad (4)$$

Z being proportional to the area under the curve of distribution, which in turn is determined by the deformation parameters. In other words, $\nu_H$ is a proportionality factor between a canonical ensemble without and with degeneracies, while $\nu_S$ relates the difference between them to the ensemble without degeneracies. Because of this relationship, the difference between the two proportionality factors has to be unity, and the following equation holds

$$\nu_S = \frac{1}{\nu_H} - 1 \quad (5)$$

and the value of $\nu_H$ is in the range $0 < \nu_H < 1$, while the value of $\nu_S$ lies in the range $0 < \nu_S < \infty$.

With these deformation parameter definitions, we can address the thermodynamic stability of an Ag molecule in the serum with reference to the binding potential of Abs. Since free energy is proportional to the logarithm of the partition function of an ensemble, proportionalities become additive values as we take the logarithm of $\nu_H$ and $\nu_S$. In terms of enthalpic and entropic contributions the above relationship can be expressed as reaction free energy of serum Ab $\Delta G_s$ against the specified Ag

$$\Delta G_s = \Delta H_s - T\Delta S_s = \Delta \mathrm{H}^\circ + \Delta \mathrm{H}^x - \mathrm{T}(\Delta \mathrm{S}^x - \Delta \mathrm{S}^\circ) \quad (6)$$

where $\Delta \mathrm{H}^x$ is excess enthalpy, $\Delta S^x$ is entropy in excess to the standard reference entropy $\Delta S^\circ$, $\Delta S_s$ is the entropic contribution to the binding energy.

$$\Delta \mathrm{G_s} = \Delta \mathrm{H}^\circ + \mathrm{RTln}(\frac{1}{\nu_\mathrm{H}}) - \mathrm{T}(\mathrm{ln}\nu_\mathrm{S}) \quad (7)$$

In terms of physical chemistry, we adjust non-ideality of the binding reaction and express deviation from ideal concentration. The former is determined by $\frac{1}{\nu_H}$, and the latter by $\nu_S$ in an equation analogous to Eq(7), as

$$\mu = \mu^\circ + \mathrm{RTln}\frac{1}{\nu_\mathrm{H}} + \mathrm{RTln}\nu_\mathrm{S} \equiv \mu^\circ + \mathrm{RTln}(\nu_\mathrm{S} + {\nu_\mathrm{S}}^2) \quad (8)$$

where $\mu$ is chemical potential, $\mu^\circ$ is reference (standard) chemical potential. This equation corresponds to the conventional expression

$$\mu = \mu^\circ + \mathrm{RTln}\gamma + \mathrm{RTln}\, {}^{\mathrm{C}}/_{\mathrm{C}^\circ} = \mu^\circ + \mathrm{RTlna} \quad (9)$$

where the activity coefficient $\gamma$ is used to adjust non-ideal activity attributable to excess enthalpy and ${}^{c}/_{c^\circ}$ is the concentration of Ab relative to the reference concentration, representing excess entropy.

Overall, by introducing parameters that relate distributions of the density of states in the funnel energy landscape we can provide a thermodynamic description of serum Abs. We can draw an analogy to the chemical potential of substances in real solutions: we dissolve Ag in a solvent of Abs and interactions between solute and solvent determine chemical potentials. While serum obviously contains molecules other than Abs (mainly water), as long as thermodynamically relevant interactions take place between Ag and Ab, we can neglect other factors. This is supported by experimental studies based on the further development of theory, as outlined in the last section.

## 4. Formation of a network of binding pathways

The proportionality relationship between standard and excess free energy, and standard and excess entropy also determines the formation of pathways of conformational changes. These pathways are natural thermodynamic networks [47–49] representing the flow of free energy. Specifically, in our system of study, these pathways are links connecting nodes of conformational states with decreasing energy. The density of states determines how these pathways merge towards the lowest energy state available in the system, in turn this confluence of pathways determines the degree distribution of nodes in a scale-free network model.

Using the exponential relationships introduced above, we can identify the properties of the scale-free directed network as the superposition of processes with different exponential rates, as reviewed by Newman [50]. The density of states g(E) as a function of energy (Eq.4) is the number of molecules within an energy range that will decrease their energy via binding and reach the bottom of the funnel (Fig.3). Therefore, it is the in-degree of a node representing the lowest energy state. The probability of states is the Boltzmann distribution (Eq.3), so we obtain the combination of exponentials as

$$p\big(g(E)\big) = p(E)\frac{dE}{dg(E)} \sim \frac{e^{-\frac{1}{\nu_H}\beta E}}{\nu_S e^{\nu_S \beta E}} = \frac{1}{\nu_S} g(E)^{-1-\frac{1}{\nu_H}/\nu_S} \quad (10)$$

where the ratio of rates therefore determines the distribution of the density of states. This power law distribution is an in-degree distribution in our landscape (Fig.3).

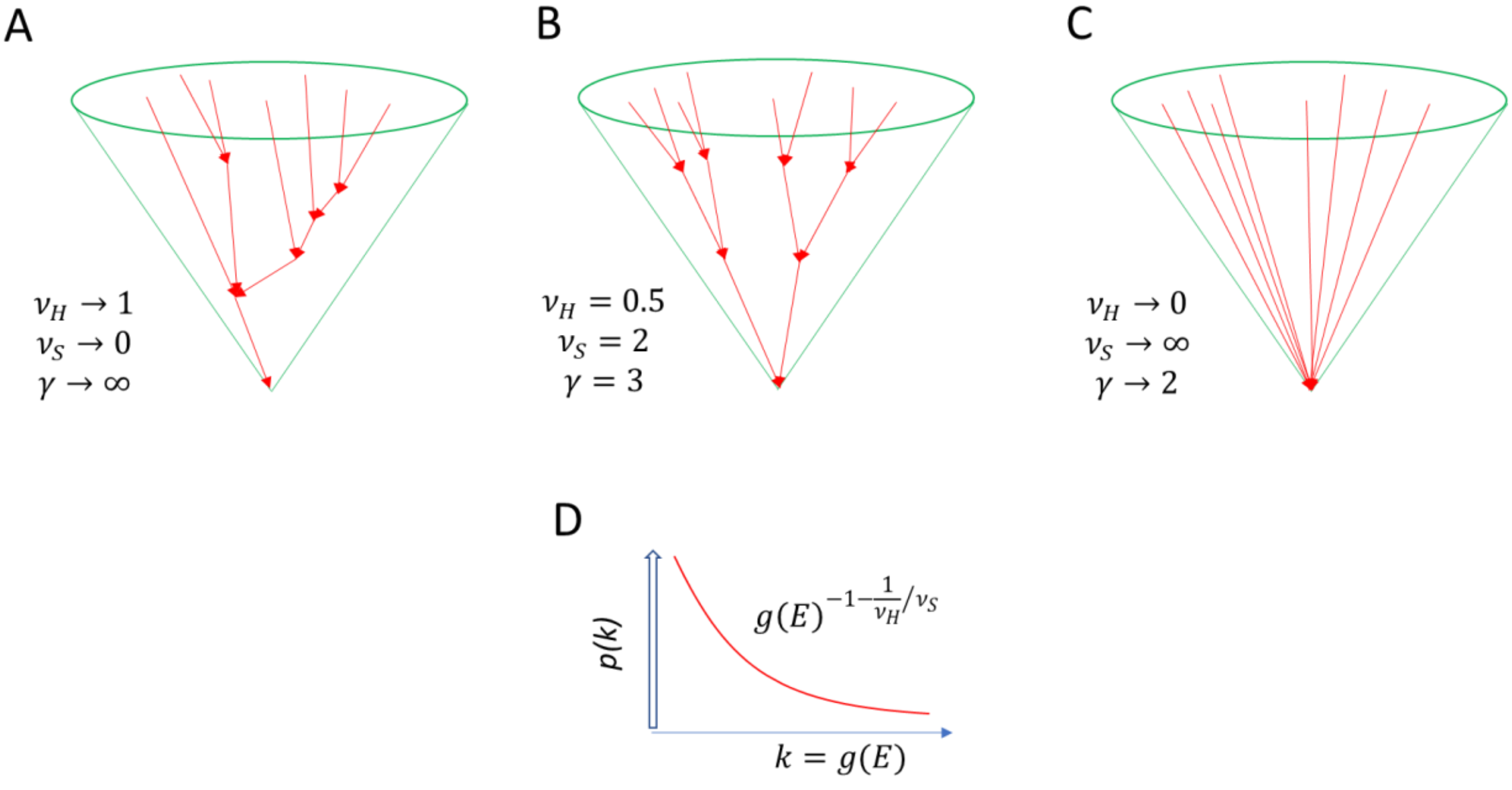

*Figure 3. Relating density of states to node degree. A-C panels show the schematic network properties of the funnel energy landscape. Scarcity of the density of states at higher energies results in networks with increasing randomness (A), while increasing density of states at high energies generates networks with greater hubs (C). The power law distribution of density of states is the in-degree distribution of the underlying network.*

As we have seen above, these rates are also ratios, which are related and convertible, so we can express the in-degree distribution exponent $\gamma$ from each as

$$\gamma = 1 + \frac{1}{\nu_H}/\nu_S = 1 + \frac{1}{1-\nu_H} = 1 + \frac{1-\nu_S}{\nu_S} \tag{11}$$

In this directed network, a system with $\nu_H = 0.5$ corresponds to the Barabási-Albert model [51,52] with a network degree distribution exponent of 3 and represents a system with an equal contribution to binding of enthalpy and excess enthalpy. The Barabási-Albert model describes a network where all the nodes increase their degree with the same rate, resulting in a hierarchic network structure (Fig.3B). That is interpreted as a balanced growth of binding energy where positive and negative entropic contributions cancel out. Our model also identifies another special value $\nu_H = 1/\varphi$, where $\varphi$ is the value of the golden ratio and in which case $\nu_S = 1/\varphi$ and $\gamma = 1 + \varphi^2$. Under these conditions excess enthalpy and excess entropy cancel out each other (see Appendix B) and the chemical potential is equal to the standard chemical potential. We speculate that this is an ideal biological state for growth since the system exerts a chemical potential towards the environment that is identical to the standard chemical potential, which is the quality, the affinity of the system. This is a balance between effector and sensor functions, allowing adaptation and evolution in the presence of a changing antigenic environment. We call this network a golden network.

## 5. Probing the super-landscape by equilibrium titration measurements

From the mathematical point of view, the Fermi-Dirac distribution FD is the complementary distribution of the logistic distribution (LD)(Appendix C). The LD is used for modeling equilibrium titration of biochemical reactions with a parametrization similar to that of FD

$$p = \frac{1}{1+e^{-(\mu-\mu^\circ)}} \tag{12}$$

where $\mu^\circ$ is the chemical potential of probe at which the reaction is halfway to completion, and p is probability of completion of reaction when $\mu$ probe potential is applied. In this parametrization $\mu$ probe potential is the logarithm of concentration of the reacting partner used for titration. This family of functions is in use in several fields of science and is also referred to as four-parameter logistic function or 4PL for immunoassays [53,54], Hill-equation for ligand binding assays [55–57], and Langmuir-equation for surface adsorption [58]. From the information theory point of view these functions are used for calibrated binary classification problems [59]: in the case of fermion particles binary classification means the occupation or emptiness of a particular energy state as a function of energy. For a biochemical reaction the function gives the probability of bound versus unbound state as chemical potential is changed. Thus, the complementarity between the FD and LD is not simply mathematical, both are concerned with particles filling energy states. While the FD distribution gives the probability of a particle filling a given energy state, at a relative distance from mean energy $\mu^\circ$, the LD gives the probability of a molecule filling its native, bound energy state, at a given applied chemical potential.

In equilibrium titration binding assays mean energy is proportional to the logarithm of equilibrium dissociation constant $K_D$; energy corresponds to Gibbs free energy and is proportional to the logarithm of concentration. While examining an Ab-Ag interaction by titrating free Ag concentration, when $\log[Ag] < \log(K_D)$ the Ab is more likely to be unbound, at $\log[Ag] = \log(K_D)$ 50% of Ab molecules are in bound state, and at $\log[Ag] > \log(K_D)$ the majority of molecules will be in bound state. Whereas the logistic function describes ideal binding and growth curves, real-life data often poorly fit such curves. The non-ideality of interactions between the probed molecules and/or the probe molecules themselves are better modeled and fitted by generalized logistic or other growth functions with more parameters [60]. The generalized logistic function introduces an asymmetry parameter to allow for non-ideality and this parameter appears as an exponent. If we assume that non-ideality stems from disproportionate contributions to binding free energy by enthalpy and entropy of binding, then we can apply measurement methods that are capable of detecting deformations in the binding curve without altering properties of the system itself [61–63]. Thus, we can use the proportionality parameters obtained above in functions describing the probing of configuration space by Ag. The enthalpic and entropic contributions to binding exerted by a heterogenous serum antibody solution on the Ag probe can be modeled by the generalized logistic distribution (Fig. 3), which we use with the parametrization of Richards [61,62,64]

$$p = \left(\frac{1}{1+\nu_H\, e^{-(\mu-\mu^{\circ})}}\right)^{\frac{1}{\nu_H}} * \left(\frac{1}{1+\nu_S\, e^{-(\mu-\mu^{\circ})}}\right)^{\frac{1}{\nu_S}} \qquad (13)$$

so as to keep the meaning of $\mu^0$ as the value at the point of inflection. Please note that the inflection point of the titration curve moves away from the inflection point of the ideal logistic curves (Fig. 3) to a position determined by the logarithm of the asymmetry parameter. The entropic contribution can be fitted by the same function where $\nu_H$ is replaced by $\nu_S$. Simultaneous titration of both thermodynamic components of binding will therefore include these two equations, as we have recently shown [64,65]. The number of available binding sites, that is the number of non-covalent bond formation per unit surface area (enthalpic contribution) is titrated by immobilizing increasing densities of antigen molecules. On the other hand, the number of available Ab molecules that contribute to funnel formation is titrated the conventional way, by serially diluting serum.

From the physico-chemical point of view, the asymmetry parameter is related to the activity coefficient, as shown in Eq.(8-9) [65]. Thus, asymmetry parameters $\nu_H$ and $\nu_S$ determine the shape of generalized logistic distribution functions of titration curves in experimental measurements.

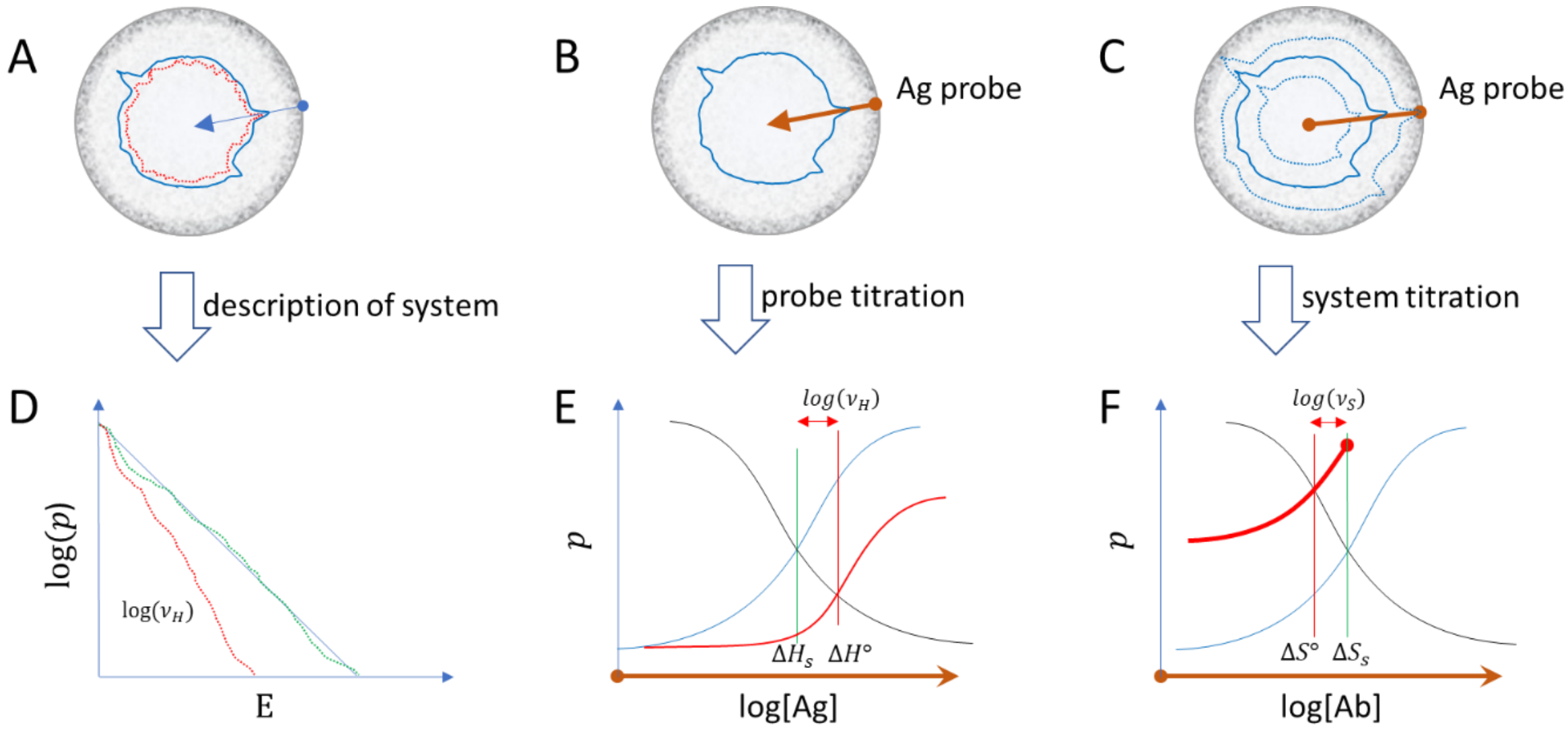

*Figure 4. Ag binding Ab landscape and models of Ag specific measurements. The ν deformation parameters define statistical distributions of the interacting system (A) and of the probed system (B, C). Binding energies in the system follow an exponential distribution (D), while immunoassays probing the system by titration reveal system potentials via generalized logistic distributions (E, F).*

## 6. Conclusions

Assays measuring the binding activity of Abs are currently based on the law of mass action. However, since no exact concentrations can be rendered to specific serum Abs, this approach is not properly applicable for serum Ab characterization. Here we argue that functions derived from the Boltzmann and the Fermi-Dirac distributions can be used for the characterization of thermodynamic potentials associated with molecular components of the adaptive immune system. The organization of interactions between components of this self-organizing system can be captured by deformation parameters, here denoted by $\nu$. These parameters are proportionality factors reflecting molecular and ensemble contributions to binding, and determining enthalpic and entropic contributions to binding free energies, as deduced from a binding energy super-landscape model. These parameters also define power law relationships often observed in complex evolving systems, hinting at an underlying scale-free, natural thermodynamic network of energy transfer. The effects of enthalpic and entropic contributions are also revealed in experimental measurements fitted with generalized logistic distributions. We propose here that this physical model identifies the key thermodynamic variables necessary for the characterization of the complex system of serum antibody interactions and lays the theoretical foundations of an experimental approach to quantitative serological assays.

**Acknowledgments**: The author wishes to thank Tamás Pfeil (Department of Applied Analysis and Computational Mathematics, Eötvös Loránd University, Budapest, Hungary) and Ágnes Kovács (Department of Biostatistics, University of Veterinary Medicine Budapest, Budapest, Hungary) for their advices and insightful discussions on the mathematics of logistic growth. Special thanks to Krisztián Papp and Zoltán Hérincs for discussions on experimental results of microarray immunoassays.

**Appendix A**

There are special cases where the chemical potential is determined by enthalpic contributions only, due to the canceling effects of entropic contributions to binding. When

$\nu_H = 0.5, \nu_S = 1$

$\mu = \mu° + RTln\frac{1}{\nu_H} + RTln\nu_S$ reduces to $\mu = \mu° + RTln\frac{1}{\nu_H}$

This case corresponds to a Barabási-Albert network model [51] with a degree distribution exponent of 3, and Ag binding from the physico-chemical perspective shows the behavior of a regular solution.

**Appendix C**

The chemical potential is determined solely by the intrinsic enthalpic binding energies when the proportionality factors are equal to the reciprocal of the golden cut, $1/\varphi$

$\nu_H = 0.618$, and $\nu_S = 0.618$

$\mu = \mu° + RTln\frac{1}{\nu_H} + RTln\nu_S$ reduces to $\mu = \mu°$

This case corresponds to a golden network [7] with degree distribution exponent $1 + \varphi^2$.

**Appendix D**

Complementarity of the Fermi-Dirac (FD) and logistic distributions (LD)

$$P(\mu) = 1 - \frac{1}{1 + e^{-\beta(E-\mu)}} = \frac{e^{-\beta(E-\mu)}}{1 + e^{-\beta(E-\mu)}} = \frac{1}{1 + e^{\beta(E-\mu)}}$$

**Conflict of interest**

The author is employed by the company Diagnosticum Zrt and is co-inventor in a patent application related to this physical model.

**Data accessibility**

Not applicable.

**Ethics statement**

Not applicable.

**Funding**

This research was not supported by any dedicated funding.